\documentclass[aps,rmp,twocolumn,secnumarabic,preprintnumbers,showpacs,showkeys]{revtex4}

\usepackage{lfmath} 
\usepackage{amsmath}
\usepackage{amsfonts}
\renewcommand{\cite}{\citep}

\begin{document}

\title{Winterberg's conjectured breaking of the superluminal quantum correlations over large distances}
\author{Eleftherios Gkioulekas}
\email{lf@mail.ucf.edu}
\affiliation{Department of Mathematics, University  of Central Florida, Orlando, FL, United States}
\begin{abstract}
We elaborate further on a hypothesis by Winterberg that  turbulent fluctuations of the zero point field may lead to a breakdown of the superluminal quantum correlations over very large distances.  A phenomenological model that was proposed by Winterberg to estimate the transition scale of the conjectured breakdown, does not lead to a  distance  that is large enough to be agreeable with recent experiments.  We consider, but rule out, the possibility of a steeper slope in the energy spectrum of the turbulent fluctuations, due to compressibility, as a possible mechanism that may lead to an increased lower-bound for the transition scale.   Instead, we argue that Winterberg overestimated the intensity of the ZPF turbulent fluctuations.  We calculate a very generous corrected lower bound for the transition  distance which is consistent with current experiments.
\end{abstract}
\preprint{submitted to \emph{Int. J. Theor. Phys.}}

\maketitle

\section{Introduction}

A compelling paradox in our current understanding of nature is the fundamental inconsistency between the non-locality of quantum mechanics and the  Lorentz invariance demanded by the theory of relativity.  This inconsistency can be concealed to a large extent because  it is possible to formulate relativistic quantum theories that predict Lorentz-invariant statistical behaviour.   Thus, from a strictly empiricist standpoint, all appears to be well since non-locality cannot be exploited to transmit information.  The inconsistency becomes more obvious when one attempts to formulate a Bohmian interpretation of quantum mechanics \cite{article:Bohm:1952,article:Bohm:1952:1,article:Bohm:1953,article:Vigier:1954,book:Hiley:1993,book:Holland:1993}.  Then it becomes necessary to introduce the quantum potential interactions which violate Lorentz invariance. There is also the problem of making the equilibrium condition $\rho=|\psi|^2$ Lorentz invariant \cite{article:Zanghi:1996}. Though resolving this latter problem is possible for the case of non-interacting but entangled particles \cite{article:Zanghi:1999}, the non-local nature of the quantum potential interaction \emph{is} unavoidable by any interpretation that agrees with experiment \cite{article:Zanghi:2004}.  Even when one decides to avoid the problem of interpretation altogether (by ignoring it) one still has to contend with the presumably   instantaneous entanglement between distant quanta in situations such as the well-known EPR two-photon experiment \cite{article:Rosen:1935,article:Bohr:1935,article:Holt:1969,article:Whitaker:2004}. 

The original Aspect experiment \cite{article:Roger:1982} confirmed the existence of  quantum entanglement  over a range as large as 10 meters by confirming the violation of Bell's inequality.  More recent experiments  have extended the range, initially to  4 km \cite{article:Owens:1994}, and subsequently to  11 km \cite{article:Gisin:1998,article:Gisin:1998:1,article:Gisin:2002} and even 50 km \cite{article:Gisin:2004}. 
   \citet{article:Winterberg:1991} observed that we cannot presume, on the basis of these experiments, that quantum correlations will indefinitely continue to hold over  larger distances. For example, it would be a bold extrapolation from a 11 km experiment to presume that quantum correlations persist over interstellar and intergalactic distances!  He also noted that the notion that the collapse of the wavefunction occurs instantaneously is probably an unjustified  idealization. It may be more reasonable to expect that quantum correlations propagate  at a finite superluminal  speed, which may be significantly larger than the speed of light, but nonetheless finite.

  If it is  true that quantum correlations break down beyond a certain ``transition''  length scale, then we are lead to the following implications: First, when we switch into  a different inertial reference frame, the transition length scale should be expected to contract over certain directions in accordance with the Lorentz transformation.  Consequently, it becomes possible to establish a unique reference frame ``at rest'', as the reference frame where the transition length scale is constant in all directions!  This then, makes the idea of the original pre-Einstein theory of relativity by Lorentz and Poincare compelling once again \cite{article:Winterberg:1987:1}.  Furthermore, if the  propagation speed of the  superluminal interactions responsible for the collapse of the wavefunction is indeed finite, that would then indicate the existence of a medium through which these interactions are  transmitted. 

Winterberg proposed the hypothesis that this medium is the field of zero-point vacuum energy, also known as the ZPF field \cite{article:Winterberg:1991,article:Winterberg:1998}. The physical reality of the ZPF field has been established experimentally by the Casimir effect \cite{article:Mostepanenko:2001}. However the underlying  physical principles that govern the ZPF field at the Planck scale are poorly understood. According to Winterberg, the simplest possible model for the ZPF field is as a ``Planck-mass  plasma'' of two coupled superfluids, one with positive mass particles and one with negative mass particles \cite{article:Winterberg:1988,article:Winterberg:1995,article:Winterberg:1997,article:Winterberg:2003,article:Winterberg:2006}. Under small perturbations, the Winterberg Planck-mass  plasma is analogous to a compressible fluid with $p \propto \rho^2$ \cite{article:Winterberg:1988}. \citet{article:Winterberg:1998}  conjectured that over large length scales this medium undergoes turbulent fluctuations which disrupt the quantum correlations at length scales where the turbulence energy spectrum exceeds the ZPF energy spectrum.

Although the physics of the ZPF field at the Planck scale is currently unknown to us, if it  is assumed that the ZPF field is  Lorentz invariant, then  it must follow the Boyer  energy spectrum   
\begin{equation}
E(k) = \hbar ck^3,
\end{equation}
which can be shown \cite{article:Boyer:1969,article:Boyer:1969:1} to be the only Lorenz invariant spectrum as long as   it is infinitely extended to arbitrarily large wavenumbers $k$.  More realistically, $k$ should be cut off at wavenumbers corresponding to the Planck length scale $r_p$, in which case Lorentz invariance would be broken only when the Planck length scale is approached.  The cutoff wavenumber defines a length, which in turn can define a rest reference frame as the one where the cut-off length is isotropic. The truncated spectrum
\begin{equation}
E(k) = \casetwo{\hbar c k^3}{k\leq 2\pi/ r_p}{0}{k>  2\pi/ r_p}
\label{eq:vaccutoff} 
\end{equation}
 is still Lorentz invariant in terms of slope and numerical coefficient  \cite{article:Gron:1987}. However,  the cutoff  scale   contracts under a Lorentz transformation like any other length. It remains a point of controversy whether or not the actual Planck length scale is the same or different for different observers.  According to the  recently proposed ``double special relativity theory'' \cite{article:Amelino-Camelia:2002} (DSR),  the Planck length scale should be the same for all observers! Under DSR, Eq. \eqref{eq:vaccutoff} would also be invariant for all inertial frames of reference! However, that would still not necessarily apply to the quantum correlations breakdown transition scale, which would still define a rest reference frame.  

Knowing the energy spectrum of the ZPF field makes it possible to estimate  the transition  length scale $\gl_0$ where quantum correlations break down by finding the transition wavenumber $k_0$ where the Boyer energy spectrum  is comparable with the energy spectrum of the ZPF turbulent fluctuations. The salient feature of this approach is that it is reliant only on our assumptions concerning  energy spectra, and does not involve further speculative modelling assumptions. \citet{article:Winterberg:1998} used the Kolmogorov-Batchelor \cite{article:Kolmogorov:1941,article:Kolmogorov:1941:1,article:Batchelor:1947} energy spectrum $E(k) \propto k^{-5/3}$  and showed that 
\begin{equation}
\gl_0 = 12 \gk^{-3/14}  \text{m}
\end{equation}
where $\gk$ represents the degree of turbulent fluctuations and ranges as $0<\gk <1$.  For $\gk=1$, we get the lower-bound length scale of  12m which is too small to be consistent with the experiments of Gisin \cite{article:Gisin:1998,article:Gisin:1998:1,article:Gisin:2002,article:Gisin:2004}. \citet{article:Winterberg:1998}  proposed  that one should use instead the evaluation $\gk = 10^{-5}$ , which is equal to the spatial temperature variation observed in the cosmic microwave radiation. This choice was thought to be plausible because ``the microwaves are refracted by the fluctuating gravitational field of the eddies'' \cite{comm:Winterberg}.  Nonetheless, a simple calculation shows that, contrary to Winterberg's claim,  this only gives us one more order of magnitude:
\begin{equation}
\gl_0 = 12 (10^{-5})^{-3/14} \text{m} \approx 120 \text{m} 
\end{equation}
This is still too small to agree with experiments. 

In the present paper we will consider the ramifications of weakening the assumption that the ZPF turbulent fluctuations are governed by the  Kolmogorov scaling $k^{-5/3}$. The idea is that a steeper slope will give a larger value for the transition length scale.  We will show that even with steeper slopes we still do not get a transition scale larger than 10 km.  Instead, we will argue that the choice $\gk=10^{-5}$ is too large because it corresponds to an unrealistically large energy dissipation rate.

The paper is organized as follows.  The energy spectrum of the turbulent fluctuations is discussed in section 2.  We recalculate the transition length scale in section 3.  The problem of guessing  the parameter $\gk$ is discussed in section 4.  Conclusions are given briefly in section 5.

\section{The spectrum of ZPF turbulence}

The underlying assumption of Kolmogorov's theory of turbulence \cite{article:Kolmogorov:1941,article:Kolmogorov:1941:1,article:Batchelor:1947} is that energy is injected into the system at small wavenumbers by random forcing, which is then cascaded to larger wavenumbers where it is being dissipated.  Between the forcing range of wavenumbers where energy comes in, and the dissipation range of wavenumbers were energy comes out, it is assumed that there is a so-called ``inertial range'' of wavenumbers through which the energy is transferred to small scales via nonlinear interactions. This theory has been confirmed both experimentally \cite{article:A.Moilliet:1962,article:Gibson:1962}, and repeatedly with numerical simulations \cite{article:Uno:2003}.  There has also been considerable progress towards formulating a statistical theory of the energy cascade \cite{lect:Procaccia:1997,article:Procaccia:2000,article:Gkioulekas:2007}.  On the other hand, it is important to note that the theory has only been investigated extensively in the context of incompressible turbulence, and it is not obvious that it is applicable for compressible turbulence.

  For the simplest case of adiabatic compressible turbulence where the pressure is dependent only on the density via the relation $p \propto \rho^{\gc}$ with $\gc$ the adiabatic constant, the Moisseev-Shivamoggi  prediction \cite{article:Yanovskii:1976,article:Shivamoggi:1992} is that the energy spectrum of the energy cascade is steeper and has slope $k^{-a}$ with $a$ given by 
\begin{equation}
a=\frac{5\gc-1}{3\gc-1}.
\end{equation}
The steeper spectral slope arises because of a distributed energy dissipation throughout the inertial range by acoustic waves. The Moisseev-Shivamoggi spectrum  reads:
\begin{equation}
E(k) = C [\gr^{\gc-1} \gee^{2\gc} c^{-2} k^{-(5\gc-1)}]^{1/(3\gc-1)}
\end{equation}
Here, $C$ is a universal constant, $\gr$ the density, $\gee$ the rate of energy injection per unit time and volume, and $k$ the wavenumber. The spectrum approaches the Kolmogorov spectrum 
\begin{equation}
E(k) = C\gr^{1/3} \gee^{2/3}  k^{-5/3}
\end{equation} when $\gc \goto\infty$ and the Kadomtsev-Petviashvili spectrum 
\begin{equation}
E(k) = C \gee c^{-1} k^{-2}
\end{equation} 
when $\gc=1$  \cite{article:Petviashvili:1973}.  Note that arbitrarily steeper slopes are possible, in principle,  if $1/3<\gc<1$, but a spectrum steeper than $k^{-3}$ would violate the locality of the nonlinear interactions that drive the energy cascade, and wouldn't really represent physically a state of fully developed turbulence.  Considering that for small fluctuations the Winterberg Planck-mass plasma behaves as a compressible fluid with $\gc=2$ \cite{article:Winterberg:1988}, the energy spectrum of compressible turbulence may be more relevant to the problem at hand. 

Let us now generalize Winterberg's calculation of the energy spectrum of ZPF turbulence for the general slope $a$. Assume that the energy spectrum of the turbulent fluctuations reads
\begin{equation}
E_{turb} (k) = A k^{-a}
\end{equation}
where $A$ is  constant and $a$ the scaling exponent. 
To estimate the constant $A$, we follow \citet{article:Winterberg:1998}, and we assume that the total energy of the turbulent fluctuations is 
\begin{equation}
\cE =\gk\gr c^2,
\label{eq:tota}
\end{equation}
 where $\gr$ is the average density of the universe and $\gk$ is a constant $0 < \gk < 1$ that measures the degree of turbulence. We estimate the density $\gr$ with the critical density $\gr_{crit}$ which separates the open and closed universe scenaria:
\begin{equation}
\gr \approx \gr_{crit} = \frac{3H^2}{8\pi G}
\end{equation}
Here $H$ is the Hubble constant and $G$ Newton's gravity constant. We also assume that the largest eddies have length scale of the order of the world radius $R \sim c/H$. It follows from these considerations that the total energy $\cE$ also satisfies
\begin{equation}
\cE = \int_{1/R}^{\infty} A k^{-a}\; dk = -\frac{A}{1-a}\frac{1}{R^{1-a}}
\label{eq:totb}
\end{equation}
and combining \eqref{eq:tota} and \eqref{eq:totb} we find $A$:
\begin{equation}
A = (a-1)\gk\gr c^2 R^{1-a}
\end{equation}

A related question, which was not addressed by Winterberg, is how the energy is injected and dissipated for the turbulent fluctuations of the field.  On a cosmological scale there is an amount of energy which is injected into the universe due to the extension of the event horizon.  This rate of energy injection represents the total available energy and it is, of course,  an upper bound to the actual rate of energy injection into the ZPF field's turbulent fluctuations, which can be very much smaller.  Following this idea, it is possible to arrive to an alternative derivation of the energy spectrum of the turbulent fluctuations of the ZPF field.  

If $R$ is the radius of the event horizon, and $\gr c^2$ the average energy density of the universe, then the rate with which the horizon expansion increases the total energy is $\gr c^2 (4\pi R^2)(dR/dt)$. To get the rate of energy injection $\gee$ we must divide this quantity with the volume of the universe. Thus, since  the radius $R$ expands with the speed of light (i.e., $dR/dt=c$), it follows that
\begin{align}
\gee &= \gr c^2 \frac{4\pi R^2 dR}{dt} \frac{1}{(4/3)\pi R^3}\\
&= 3\gr c^2 R^{-1}c = 3\gr c^3 R^{-1}
\end{align}
Substituting $\gee$ to the Moisseev-Shivamoggi spectrum, and using the speed of light for $c$, yields:
\begin{align}
E(k) &= C [\gr^{\gc-1} (3\gr c^3 R^{-1})^{2\gc} c^{-2} k^{-(5\gc-1)}]^{1/(3\gc-1)}\\
 &= C [\gr^{3\gc-1} 3^{2\gc} R^{-2\gc} c^{6\gc-2} k^{-(5\gc-1)}]^{1/(3\gc-1)}\\
 &\approx \gr c^2 R^{-2\gc/(3\gc-1)} k^{-(5\gc-1)/(3\gc-1)}
\end{align}
It is easy to see that solving $a=(5\gc-1)/(3\gc-1)$ for $\gc$ gives $\gc=(1-a)/(5-3a)$ and that it follows that $2\gc/(3\gc-1) = a-1$. Thus, we can write $E(k)$ in terms of $a$ as follows:
\begin{equation}
E(k) \approx \gr c^2 R^{1-a} k^{-a}
\end{equation}
Here we have disregarded the numerical coefficients. We see that we essentially recover Winterberg's spectrum for the case $\gk=1$. However, if instead of $\gr c^2$ we use $\gk\gr c^2$ for the universe energy density, we do not obtain exactly the same dependence on $\gk$ as we would from the generalization of Winterberg's argument! So there is some ambiguity on how one should define the degree  $\gk$ of the turbulent fluctuations. 

\section{Estimating the transition scale}

We will now show that a steeper slope for the energy spectrum of the turbulent fluctuations gives a larger transition length scale.  However, we will see that there is no case that gives a large enough length scale to be consistent with Gisin's  experiments  \cite{article:Gisin:1998,article:Gisin:1998:1,article:Gisin:2002,article:Gisin:2004}, unless, a smaller value of $\gk$ is chosen.

The energy spectrum of the zero-point vacuum energy is given by
\begin{equation}
E_{vac} (k) = \hbar ck^3
\end{equation}
The transition  wavenumber $k_0$ where we may expect quantum correlations to break can be estinated, in the sense of an upper-bound, by matching the two energy spectra $E_{vac} (k)$ and $E_{turb} (k)= A k^{-a}$, which yields
\begin{equation}
k \approx \fracp{A}{\hbar c}^{\frac{1}{3+a}} \equiv k_0 
\end{equation}
Here we deviate slightly from \citet{article:Winterberg:1998} who compared the cumulative spectra instead. The advantage of this approach is that we don't need to make the assumption that these power laws hold at all scales. Instead, it is sufficient that they hold around the intersection wavenumber. 

Using the evaluation $A = (a-1)\gk\gr c^2 R^{1-a}$ the transition wavenumber $k_0$ reads:
\begin{align}
k_0 &= \fracp{(a-1)\gk\gr c^2 R^{1-a}}{\hbar c}^{\frac{1}{3+a}} \\
&= \left[ (a-1) \frac{\gk c}{\hbar} \fracp{3H^2}{8\pi G}\fracp{c}{H}^{1-a} \right]^{\frac{1}{3+a}} \\
&= \left[ \frac{3(a-1)\gk c^2 H}{8\pi G\hbar} \fracp{H}{c}^a \right]^{\frac{1}{3+a}}
\end{align}
This result generalizes equation (10) in \citet{article:Winterberg:1998}. The corresponding cross-over length scale reads
\begin{equation}
\gl_0= \frac{2\pi}{k_0} = 2\pi\left[ \frac{8\pi G\hbar}{3(a-1)\gk c^2 H} \fracp{c}{H}^a \right]^{\frac{1}{3+a}}
\end{equation}
On a decimal logarithmic scale, the order of magnitude of $\gl_0$ is 
\begin{widetext}
\begin{equation}
\log \gl_0 = \log (2\pi)-\frac{1}{3+a} \left[ \log  \frac{3 c^2 H}{8\pi G\hbar} + \log\gk + \log (a-1) + a \log \fracp{H}{c} \right].
\end{equation}

Let $\gs_{\hbar}$, $\gs_{G}$, $\gs_{c}$, and $\gs_{H}$ be the measurement errors of the corresponding constants  $\hbar$, $G$, $c$, and $H$, and assume for simplicity that these errors are statistically uncorrelated. The propagated statistical uncertainty is given by:
\begin{align}
\gs_{\gl_0}^2 &= \left[ \pderiv{\gl_0}{G} \gs_{G} \right]^2 + \left[ \pderiv{\gl_0}{c} \gs_{c} \right]^2 + \left[ \pderiv{\gl_0}{\hbar} \gs_{\hbar} \right]^2 + \left[ \pderiv{\gl_0}{H} \gs_{H} \right]^2  \\
 &= \left[ \frac{1}{3+a} \frac{\gl_0\gs_{G}}{G}  \right]^2 + \left[ \frac{a-2}{3+a} \frac{\gl_0\gs_{c}}{c}  \right]^2 + \left[ \frac{1}{3+a} \frac{\gl_0\gs_{\hbar}}{\hbar}  \right]^2 + \left[ \frac{1}{3+a} \frac{\gl_0\gs_{H}}{H}   \right]^2 
\end{align}
Consequently, in terms of the relative standard uncertainties $e_{c} = \gs_{c}/c$, $e_{G} = \gs_{G}/G$, $e_{\hbar} = \gs_{\hbar}/\hbar$,  $e_{H} = \gs_{H}/H$, and $e_{\gk} = \gs_{\gk}/\gk$ we obtain
\begin{equation}
\fracp{\gs_{\gl_0}}{\gl_0}^2 = \frac{e_{G}^2+ (a-2)^2 e_{c}^2 + e_{\hbar}^2 + (a+1)^2 e_{H}^2 + e_{\gk}^2 }{(3+a)^2}
\end{equation}
\end{widetext}
The 2006 CODATA recommended values for the universal constants are: $c=2997924583 \;\text{m}/\text{s}$,  $G= 6.67428\times 10^{-11} \;\text{N}\cdot \text{m}^2 \text{kg}^{-2}$, $\hbar = 1.054571628\times 10^{-34} \;\text{J}\cdot \text{s}$. The corresponding relative standard uncertainties are: $e_{c} = 0$, $e_{G} = 10^{-4}$, $e_{\hbar} = 5\cdot 10^{-5}$. Furthermore, the recent measurement \cite{article:Dawson:2006} of the Hubble constant  by NASA's Chandra X-ray Observatory gives   $H= 77 \;\text{km}\; \text{s}^{-1} \text{Mpc}^{-1}$ (with $1 \text{Mpc} = 3.26 \cdot 10^6$ lightyears, we get $H= 2.49\cdot 10^{-18} \;\text{s}^{-1}$). The uncertainty of $H$ is $e_{H} = 0.15$, which is by far the dominant contribution to $\gs_{\gl_0}$. We may thus estimate $\gs_{\gl_0}$ practically via:
\begin{equation}
\fracp{\gs_{\gl_0}}{\gl_0} \approx \frac{(a+1) e_{H}}{(a+3)}
\end{equation}
From these evaluations we also obtain the following empirical formula for the transition wavelength $\gl_0$ 
\begin{widetext}
\begin{equation}
\log\gl_0 = \log (2\pi)-\frac{98.045658-60.049519a-\log\gk -\log (a-1)}{a+3}
\end{equation}
\end{widetext}
The relevant slopes $a$ are: Kolmogorov scaling with intermittency corrections (i.e. $a=1.7$), compressible scaling with $\gc=2$ (i.e. $a=1.8$). It is also worthwhile to consider the extreme case $a=2$ which corresponds to the shock-dominated Kadomtsev-Petviashvili spectrum. For $\gk=1$, we get the following transition scales: $a = 1.7$ gives $\gl_0 = (16\pm 1) \text{m}$, $a = 1.8$ gives $\gl_0 = (53 \pm 4) \text{m}$, and $a = 2$ gives $\gl_0 = (517 \pm 46) \text{m}$. The transition scales are increased for $\gk=10^{-5}$ (Winterberg's choice) as follows: $a = 1.7$ gives $\gl_0 = ( 185\pm 15 ) \text{m}$, $a = 1.8$ gives $\gl_0 =  (587 \pm 51) \text{m}$, and $a = 2$ gives $\gl_0 = (5172 \pm 465) \text{m}$.
In all cases the transition scale is less than $11 \text{km}$. This suggests that Winterberg's choice for $\gk$ is probably too large.

\section{What $\gk$ is reasonable?}

To get a good sense of what $\gk$ is reasonable, we shall calculate the energy dissipation rate $\gee$ in terms of $\gk$. From the Moisseev-Shivamoggi spectrum we see that $A$ reads:
\begin{equation}
A = [\gr^{\gc-1} \gee^{2\gc} c^{-2}]^{1/(3\gc-1)}.
\end{equation}
Here we have assumed that the Kolmogorov constant is unity, since we shall be doing only order of magnitude calculations. Solving for $\gee$ we obtain
\begin{align}
\gee &= \fracb{A^{3\gc-1}}{\gr^{\gc-1} c^{-2}}^{1/2\gc}= \fracb{[(a-1)\gk\gr c^2 R^{1-a}]^{3\gc-1}}{\gr^{\gc-1} c^{-2}}^{1/2\gc}\\
&= [[(a-1)\gk R^{1-a}]^{3\gc-1} \gr^{2\gc} c^{6\gc}]^{1/2\gc}\\
&= \gr c^3 [(a-1)\gk R^{1-a}]^{(3\gc-1)/(2\gc)}
\end{align}
This expression can be simplified further because the identity $2\gc/(3\gc-1) = a-1$ implies that
\begin{equation}
\frac{(1-a)(3\gc-1)}{2\gc}= -1.
\end{equation}
Thus the energy dissipation rate can be rewritten as
\begin{equation}
\gee =  \gr c^3 R^{-1} [(a-1)\gk]^{1/(a-1)}.
\end{equation}

To get some physical understanding,  compare the amount of energy dissipated over the entire universe under the rate $\gee$ during time $t$,  with the energy that would be released by the annihilation of  $N$  solar masses $M$. Then, $N M c^2 = \gee R^3 t $, and it follows that
\begin{align}
N &= \frac{\gee R^3 t}{M c^2} = \frac{\gr c^3 R^{-1} [(a-1)\gk]^{1/(a-1)} R^3 t}{M c^2} \\
&= \frac{\gr c R^2t}{M} [(a-1)\gk]^{1/(a-1)} = N_0 [(a-1)\gk]^{1/(a-1)} 
\end{align}
where $N_0 = \gr c R^2 t/M$ is a dimensionless  constant. The reverse relation between $\gk$ and $N$ is:
\begin{equation}
\gk = \frac{1}{a-1}\fracp{N}{N_0}^{a-1}
\end{equation}
Using $M=1.98 \cdot 10^{30} \text{kg}$ (the mass of our Sun) and $t= 1 \text{day}$, yields $N_0 = 10^{57}$. The value $\gk=10^{-5}$ that was proposed by Winterberg gives $N\approx10^{49}$ for $a=1.7$.  To get some sense of what this means, consider the rescaled count 
\begin{equation}
N_s = N\fracp{\ell}{R}^3 = N \fracp{\ell H}{c}^3, 
\end{equation}
 with $\ell = 8 \text{ lightminutes}$,  which is approximately the Earth-Sun distance.  We find $N_s \approx10^{5}$ .  This means that every day, within the neighborhood of  Earth's planetary orbit,  the  amount of the energy  dissipated should be equal to the energy that would be released by the complete annihilation of $10^{5}$ solar masses! For $a=1.8$  this count increases to $N_s=10^6$ solar masses per day.  Such an extraordinary amount of energy should have been somewhat conspicuous to all life on Earth!  This is why I believe that $\gk$ should be chosen to be much smaller.

As we have mentioned previously, turbulence requires both a mechanism for injecting energy into the system, and a mechanism for dissipating the energy at large wavenumbers.  The question of finding a reasonable choice for the variable $\gk$, is equivalent to the question of deciding on an upper bound for the dissipation rate of the turbulent fluctuations of the ZPF field.  Naturally, the underlying problem is understanding the  physical mechanism  responsible for dissipating the energy cascade of ZPF turbulence in the first place! 

Rather than speculate on the particulars of the dissipation mechanism, we can simply assume that the rate of energy output from the Sun is an extremely generous upper bound for the rate of energy dissipation over a spherical volume around the Sun that reaches the Earth.  Since the Sun has an overall lifetime of approximately 10 billion years (which is approximately $10^{12}$ days), we can further overestimate the rate of energy output by choosing $N_s=10^{-12}$ solar masses per day. The corresponding $\gk$ is $a$ dependent, and it is given by
\begin{equation}
\gk = \frac{1}{a-1}\left(\frac{N_s}{N_0}\fracp{c}{\ell H}^3 \right)^{a-1}
\end{equation}
 For $a=1.7$ , this gives $\gk=9\times 10^{-18}$ , from which we get for the transition scale $\gl_0 = 67\pm 5$ km.  The situation is improved if we choose $a=1.8$ .  Then we get $\gk=2 \times10^{-20}$ which gives $\gl_0 = 630\pm 55$ km.  These numbers are extremely generous underestimates for any reasonable evaluation of the transition scale and they are both agreeable with Gisin's experiments \cite{article:Gisin:1998,article:Gisin:1998:1,article:Gisin:2002,article:Gisin:2004}. The actual transition scale probably exceeds these by some orders of magnitude and could be as large as, for example, the Earth-Moon distance. 

\section{Conclusion and Discussion}

In the present paper I have shown that the model of the ZPF turbulent fluctuations proposed by \citet{article:Winterberg:1998} to account for the conjectured breakdown of the quantum correlations is plausible in the sense that it does not disagree with current experiments.  However, it is still unsatisfactory  to not have a clear understanding of where the turbulence gets the energy, how the energy gets dissipated, and where it ends up after it is dissipated.  This aspect of the model requires further elaboration.

The strange behavior of quantum-mechanical systems that involves entanglement over large distances is a very tantalizing mystery.  It has led most physicists to the very extreme position of denying the existence of an objective reality underlying quantum mechanics (the Copenhagen interpretation). This  prompted Einstein to comment in a memorable way on the non-existence of  God's gambling addiction!  A more mild position along the same lines, which is still nonetheless a  partial denial of objective reality, is the ``relational interpretation'' \cite{article:Laudisa:2001,article:Filk:2006,article:Rovelli:2007}.  The idea here is to deny only as much of  objective reality as is necessary to make the existing problems go away. 

 Part of the mystery is that we don't really understand what the wavefunction really is.  \citet{article:Winterberg:1991} assumes that the wavefunction is a genuine physical field that really collapses. In the same paper he also reviews the early literature on the subject.  It is hard to accept this viewpoint and not expect the collapse to propagate at a finite speed, or to not be disrupted by possible noise in the mechanism that propagates it. From the standpoint of the Bohmian interpretation, the wavefunction of the combined physical system and the measuring apparatus never really collapses!  Nonetheless, even in the Bohmian interpretation, one \emph{models}  (instead of deriving from first principles) the Hamiltonian governing the interaction between system and apparatus during measurement. A model is just a model and there is no need for it to be exact. Furthermore, one can expect a breakdown in quantum correlations if there is a small amount of noise, presumably  from subquantum processes,  in the guidance condition that determines the particle velocities from the wave function. 

There is also some controversy over the experiments that have convinced us of the reality of quantum correlations in the first place.  It is believed by some that these experiments are susceptible to certain ``loopholes'' (the locality loophole and the detection loophole) that prevent them from being conclusive \cite{article:Santos:1990,article:Santos:1991,article:Santos:1992,article:Santos:2004}.  There has been some interest in developing experiments that tried to close the loopholes \cite{article:Zeilinger:1986,article:Zeilinger:1998,article:Aspect:1999,article:Grangier:2001,article:Wineland:2001,article:Aspect:2007,article:Zeilinger:2007}.  Nonetheless, the controversy continues \cite{article:Santos:2007}. It may turn out that, over increasing distances, these experiments begin to fail gradually. The ``loopholes'' could then represent modes of failure that are ``irrelevant'' at short distances but become increasingly relevant over larger distances.   
Even if the particulars of the model proposed by Winterberg  turn out to be wrong, this underlying issue of understanding the possible role of distance with respect to entanglement  remains. 

\begin{acknowledgements}
The author would like to thank Dr. Friedwardt Winterberg for his e-mail correspondance. 
\end{acknowledgements}

\bibliography{references,references-submit,references-winterberg}
\bibliographystyle{apsrmp}

\end{document}